\documentclass[prc,twocolumn,superscriptaddress,showpacs]{revtex4}
\usepackage{graphicx,amsmath,amssymb,bm}

\newcommand{\la}{\Lambda}
\newcommand{\vlowk}{V_{{\rm low}\,k}}
\newcommand{\fmi}{\, \text{fm}^{-1}}
\newcommand{\mev}{\, \text{MeV}}
\newcommand{\kev}{\, \text{keV}}
\newcommand{\hw}{\hbar \omega}

\begin{document}

\title{Benchmark calculations for $^3$H, $^4$He, $^{16}$O and $^{40} 
$Ca with {\it ab-initio} coupled-cluster theory}

\author{G.~Hagen}
\affiliation{Physics Division, Oak Ridge National Laboratory,
Oak Ridge, TN 37831, USA}
\affiliation{Department of Physics and Astronomy, University of  
Tennessee, Knoxville, TN 37996, USA}
\affiliation{Center of Mathematics for Applications, University of  
Oslo, N-0316 Oslo, Norway}
\author{D.J.~Dean}
\affiliation{Physics Division, Oak Ridge National Laboratory,
Oak Ridge, TN 37831, USA}
\author{M.~Hjorth-Jensen}
\affiliation{Department of Physics and Center of Mathematics for  
Applications, University of Oslo, N-0316 Oslo, Norway}
\author{T.~Papenbrock}
\affiliation{Physics Division, Oak Ridge National Laboratory,
Oak Ridge, TN 37831, USA}
\affiliation{Department of Physics and Astronomy, University of  
Tennessee, Knoxville, TN 37996, USA}
\author{A.~Schwenk}
\affiliation{TRIUMF, 4004 Wesbrook Mall, Vancouver, BC, Canada, V6T 2A3}

\begin{abstract}
We present {\it ab-initio} calculations for $^3$H, $^4$He, $^{16}$O,
and $^{40}$Ca based on two-nucleon low-momentum interactions $\vlowk$
within coupled-cluster theory. For $^3$H and $^4$He, our results are
within $70 \kev$ and $10 \kev$ of the corresponding Faddeev and
Faddeev-Yakubovsky energies. We study in detail the convergence with
respect to the size of the model space and the single-particle
basis. For the heavier nuclei, we report practically converged binding
energies and compare with other approaches.
\end{abstract}


\maketitle

\section{Introduction}

{\it Ab-initio} few- and many-body methods have been used with great
success to explore the structure of light nuclei based on microscopic
two- and three-nucleon interactions. For nuclei with $A \lesssim 12$
nucleons, several techniques provide practically exact solutions to
the nuclear many-body problem and have been benchmarked to agree within
numerical uncertainties for the $^4$He ground-state energy and radius
obtained from the nucleon-nucleon (NN) Argonne $v_8$ potential~\cite{Kam01}.
These methods include Faddeev-Yakubovsky equations~\cite{Nog00}, 
variational approaches~\cite{Var95,Hiy03,Viv04,Viv05}, the Green's
function Monte Carlo (GFMC) method~\cite{Piep01}, the No-Core Shell 
Model (NCSM)~\cite{Nav00,Nav07}, and the effective interaction 
hyperspherical harmonics method~\cite{Bar99,Gaz05}. Many of
the above approaches are, however, restricted to the lightest nuclei.

In recent years, the coupled-cluster method~\cite{Coe58,Coe60} was
reintroduced in nuclear physics as a tool for {\it ab-initio} nuclear
structure calculations~\cite{Hei99,Mi00,Dean04,Kow04,Wlo05,Gour06,%
Hag06,Hag07}. Coupled-cluster theory is size-extensive and scales
rather gently with an increasing number of nucleons and with the size
of the model space, and therefore has the potential to extend the
reach to medium-mass nuclei.  In this paper, we address the question
of whether the coupled-cluster method is as precise for few-body
systems as the well-established methods.  Several findings suggest
that this is indeed the case. Mihaila and Heisenberg performed a
microscopic calculation of the electron scattering structure function
for $^{16}$O and found excellent agreement with experimental data.
Their calculations are based on a particle-hole energy expansion of
the cluster operator. More recent
applications~\cite{Dean04,Kow04,Wlo05,Gour06,Hag06,Hag07} follow the
``standard'' approach of coupled-cluster calculations from quantum
chemistry~\cite{Bar07,Pal99,Craw00,Pie04}. In these calculations,
several results for helium isotopes~\cite{Kow04,Hag06} were found to
be in good agreement with exact diagonalizations in sufficiently small
model spaces and with corresponding renormalized
interactions. However, except in the recent study of three-nucleon
forces (3NF) in coupled-cluster theory~\cite{Hag07}, this approach has
never been compared in detail to well-established few-body methods in
the larger model spaces that are needed for convergence with modern NN
interactions. It is the purpose of this work to fill this gap in
nuclear physics, and to place the coupled-cluster method in the group
of {\it ab-initio} approaches.

This paper is organized as follows. In Section~\ref{backgr}, we begin
with a brief discussion of the coupled-cluster method, of the
low-momentum interactions and the employed basis spaces. Our main
results for $^3$H and $^4$He are presented in Section~\ref{res1}, and
for $^{16}$O and $^{40}$Ca in Section~\ref{res2}. We conclude with a
summary in Section~\ref{sum}.

\section{Method, interactions, and model spaces}
\label{backgr}

\subsection{Coupled-cluster method}
\label{cc}

Coupled-cluster theory was invented by Coester and K\"ummel almost
fifty years ago \cite{Coe58,Coe60}. During the 1970s, this approach
was further developed and found many applications in nuclear physics.
The review by the Bochum group~\cite{Kuem78}
summarizes the status of the field in 1978. From there on,
applications in nuclear physics were more of a sporadic 
nature~\cite{Guardiola}.
This was most probably due to the difficulty of hard NN interactions
and their strong short-range repulsion and short-range tensor force.
Mihaila and Heisenberg employed coupled-cluster theory in the late 
1990s~\cite{Hei99}. Their work culminated in the precise computation 
of the electron scattering form factor for $^{16}$O based on the 
Argonne $v_{18}$ potential combined with leading contributions 
from 3NF~\cite{Mi00}.

Parallel to the field of nuclear physics, coupled-cluster theory saw
its own career in {\it ab-initio} quantum chemistry. After the
pioneering works by {{\v C}{\'\i}{\v z}ek}~\cite{Ciz66,Ciz69}, the
theory has become one of the main workhorses in quantum
chemistry~\cite{Bar07,Pal99,Craw00,Pie04}. The sheer number of
applications and developments in that field {\it de-facto} established
a ``standard'' or ``canonical'' way for how the method is being used
to solve quantum many-body problems. Reference~\cite{Bar07} gives a
summary of state-of-the-art coupled-cluster calculations in  
quantum chemistry.

Recently, coupled-cluster theory has seen a renaissance in nuclear
physics starting with the calculations of Ref.~\cite{Dean04}. This 
approach differs from the one by Mihaila and Heisenberg as it employs 
coupled-cluster theory in the spirit of quantum chemistry and uses
softer interactions. So far, the present approach has employed
$G$ matrices for the description of ground and excited states in
$^4$He~\cite{Kow04} and $^{16}$O~\cite{Wlo05}, and for nuclei in the
vicinity of $^{16}$O~\cite{Gour06}. The most recent calculations are
based directly on low-momentum interactions $\vlowk$~\cite{Vlowk1,Vlowk2},
and the method has been developed to describe weakly bound and 
unbound helium isotopes within a Gamow-Hartree-Fock basis~\cite{Hag06} and
to include 3NF~\cite{Hag07}.

Within coupled-cluster theory, the ground state of a mass $A$ nucleus
is written as
\begin{equation}
\label{cc_ansatz}
|\psi\rangle = e^{\hat{T}} |\phi\rangle \,,
\end{equation}
where $|\phi\rangle=\prod_{i=1}^A \hat{a}_i^\dagger|0\rangle$
is a single-particle product state and
\begin{equation}
\label{T}
\hat{T} = \hat{T}_1 + \hat{T}_2 + \ldots + \hat{T}_A
\end{equation}
is a particle-hole ($p$-$h$) excitation operator with
\begin{equation}
\hat{T}_k =
\frac{1}{(k!)^2} \sum_{i_1,\ldots,i_k; a_1,\ldots,a_k} t_{i_1\ldots
i_k}^{a_1\ldots a_k}
\hat{a}^\dagger_{a_1}\ldots\hat{a}^\dagger_{a_k}
\hat{a}_{i_k}\ldots\hat{a}_{i_1} \,.
\end{equation}
Here and in the following, $i, j, k,\ldots$ label occupied 
single-particle orbitals (as defined by the product state $|\phi\rangle$)
while $a, b, c,\ldots$ refer to unoccupied orbitals.

We take the reference state $|\phi\rangle$ as our vacuum state
and normal-order the Hamiltonian $\hat{H}$ with respect to this
state. In practice, we restrict ourselves to the truncation $\hat{T}= 
\hat{T}_1+\hat{T}_2$. This is the CCSD approximation, and the 
coupled-cluster equations are given by
\begin{eqnarray}
\label{ccsd}
E &=& \langle \phi | \overline{H} | \phi\rangle \,, \\
\label{ccsd1}
0 &=& \langle \phi_i^a | \overline{H} | \phi\rangle \,, \\
\label{ccsd2}
0 &=& \langle \phi_{ij}^{ab} | \overline{H} | \phi\rangle \,.
\end{eqnarray}
Here $|\phi_{i_1\ldots i_n}^{a_1\ldots a_n}\rangle = 
\hat{a}_{a_n}^\dagger\ldots \hat{a}_{a_1}^\dagger 
\hat{a}_{i_1}\ldots \hat{a}_{i_n}|\phi\rangle$ is a $np$-$nh$ 
excitation of the reference state $|\phi\rangle$, and 
\begin{equation}
\label{hsim}
\overline{H} = e^{-\hat{T}} \hat{H} e^{\hat{T}} 
= \bigl( \hat{H} e^{\hat{T}} \bigr)_c
\end{equation}
is the similarity-transformed Hamiltonian (note that $\overline{H}$
is non-Hermitian). 
The last expression on the right-hand side of Eq.~(\ref{hsim}) 
indicates that only fully connected 
diagrams contribute to the construction. 
The CCSD Eqs.~(\ref{ccsd1}) and (\ref{ccsd2})
determine the amplitudes $t_i^a$ and $t_{ij}^{ab}$ of the $1p$-$1h$ and
the $2p$-$2h$ excitation cluster operators, respectively. Once these
nonlinear equations are solved, the amplitudes can be inserted into 
Eq.~(\ref{ccsd}) to determine the ground-state energy.

We remind the reader that an exact solution of the many-body problem
would require us to employ the full excitation operator
Eq.~(\ref{T}). Such a calculation is as expensive as a full
diagonalization of the Hamiltonian, and therefore impossible for
medium-mass nuclei. CCSD is very efficient in the sense that it is a
highly accurate approximation with the investment of a modest
numerical effort that scales as $O(n_o^2n_u^4)$ with the number $n_o$
of occupied and the number $n_u$ of unoccupied single-particle
orbitals, respectively. The inclusion of the $3p$-$3h$ cluster operator
$\hat{T}_3$ would further increase the accuracy of the
method. However, such CCSDT calculations come at the expense
$O(n_o^3n_u^5)$ and, at present, are already prohibitively expensive
compared to CCSD.  For this reason, there is need for more approximate
treatments of the full triples equations.

There are various approximations to the full
CCSDT equations, and the most popular of these schemes is the CCSD(T)
approach \cite{Deegan}. 
CCSD(T) includes diagrams at the CCSDT level that
appear up to fifth order in perturbation theory. It is a
non-iterative approach since typically converged singles and doubles
excitation amplitudes are used in the calculation of the  triples
energy correction. The CCSD(T) approximation is
relatively inexpensive compared to CCSDT; no storage of triples
amplitudes is required and the computational cost is a non-iterative
$O(n_o^3n_u^4)$ step. There is also a family of iterative triples
correction schemes known as CCSDT-$n$~\cite{ccsdt-n}.
Their derivation is based on perturbation theory arguments,
\begin{eqnarray}
\nonumber
&\text{CCSDT-1} \quad
&0 = \langle \phi_{ijk}^{abc} | 
\bigl( \hat{F} \hat{T}_3 + \hat{H} \hat{T}_2 \bigr)_c |
\phi \rangle \,,\\
\nonumber
&\text{CCSDT-2} \quad
&0 = \langle \phi_{ijk}^{abc} |
\bigl( \hat{F} \hat{T}_3 + \hat{H} \hat{T}_2
+ \hat{H} \hat{T}_2^2/2 \bigr)_c |
\phi \rangle \,,\\
\nonumber
&\text{CCSDT-3} \quad
&0 = \langle \phi_{ijk}^{abc} |
\bigl( \hat{F} \hat{T}_3 + \hat{H} e^{ \hat{T}_1 + \hat{T}_2} \bigr)_c |
\phi \rangle \,,\\
&\text{CCSDT} \quad
&0 = \langle \phi_{ijk}^{abc} |
\bigl( \hat{H} e^{ \hat{T}_1 + \hat{T}_2 + \hat{T}_3} \bigr)_c |
\phi \rangle \,.
\end{eqnarray}
Here, $\hat{F}$ denotes the Fock operator (the one-body operator that 
results from the normal ordering of the Hamiltonian).
All these approaches require the storage of the full triples
amplitudes $t_{ijk}^{abc}$ and are therefore computationally
considerably more expensive than the CCSD(T) approach. However,
for cases where the CCSD(T) scheme breaks down, one expects
the CCSDT-$n$ approaches to perform better. The latter approaches treat
the triples corrections self-consistently and also involve the 
corrections 
\begin{eqnarray}
\langle \phi_{i}^{a} | 
\bigl( \hat{V} \hat{T}_3 \bigr)_c | \phi \rangle \,,
\label{ccsdt11} \\
\langle \phi_{ij}^{ab} | 
\bigl( \hat{F} \hat{T}_3 + \hat{V} \hat{T}_3 
+ \hat{V} \hat{T}_3 \hat{T}_1 \bigr)_c | \phi \rangle \,,
\label{ccsdt12}
\end{eqnarray}
to Eqs.~(\ref{ccsd1}) and~(\ref{ccsd2}), respectively. These corrections
thus modify the values of the amplitudes $t_i^a$ and $t_{ij}^{ab}$. 

\subsection{Low-momentum interactions and model spaces}
\label{nn}

Nuclear interactions depend on the resolution scale at which
details are probed and resolved.
This resolution scale dependence is similar to scale and scheme 
dependences in parton distribution functions. As a result, nuclear 
interactions are defined by an effective theory for NN, 3N, and 
many-nucleon interactions and corresponding effective operators,
\begin{equation}
\hat{V} = V_{\rm NN}(\la) + V_{\rm 3N}(\la) + \ldots \,,
\end{equation}
where the momentum cutoff $\la$ denotes the resolution scale.
Conventional nuclear forces are ``hard'' in the sense that they
have large cutoffs that complicate few- and many-body calculations.
These difficulties arise from high momenta and associated strong 
short-range repulsion and short-range tensor forces, which lead
to slow convergence with increasing basis size and requires resummations in
practice.

Low-momentum interactions $\vlowk$ with variable momentum cutoffs
show great promise for nuclei~\cite{Vlowk1,Vlowk2,Vlowk3N,Vlowknuc,%
VlowkSM1,VlowkSM2,variational1,variational2}. 
Changing the cutoff leaves 
low-energy NN observables unchanged by construction, but shifts
contributions between the potential and the sums over intermediate
states in loop integrals.  These shifts can weaken or largely
eliminate sources of nonperturbative behavior such as strong
short-range repulsion and the tensor force~\cite{Vlowknuc,Born}. An
additional advantage is that the corresponding 3N interactions become
perturbative at lower cutoffs~\cite{Vlowk3N} and are thus tractable in
coupled-cluster theory~\cite{Hag07}. The renormalization group (RG)
evolution is implemented by coupled RG equations in momentum
space~\cite{VlowkRG} or by an equivalent Lee-Suzuki
transformation~\cite{LS1,LS2}.

The evolution to low-momentum interactions $\vlowk$ weakens
off-diagonal coupling and decouples the low-energy physics from
high-momentum details~\cite{SRG,decoupling}. As a result, few- and
many-body calculations converge more rapidly for lower cutoffs,
which is important for extending {\it ab-initio} approaches to
heavier systems. Finally, the cutoff variation can provide estimates
for theoretical uncertainties, which will be left to future work.
In this paper, we use a sharp cutoff $\la = 1.9 \fmi$ for the
$^3$H and $^4$He calculations, and $\vlowk$ is derived from the 
Argonne $v_{18}$ potential~\cite{AV18} in order to benchmark
against the Faddeev and Faddeev-Yakubovsky results~\cite{Vlowk3N}.
For $^{16}$O and $^{40}$Ca, we use a cutoff $\la = 2.1 \fmi$ and
compare to the importance-truncated NCSM study~\cite{Roth07}.

Coupled-cluster theory is employed in a single-particle basis, and
we use a model space consisting of spherical harmonic-oscillator
states. The basis parameters are the number of orbitals and 
the oscillator frequency $\hw$. Our largest model spaces include
about $10^3$ single-particle orbitals.

\section{Results for $^3$H and $^4$He}
\label{res1}

\begin{figure}[t]
\includegraphics[width=0.45\textwidth,clip=]{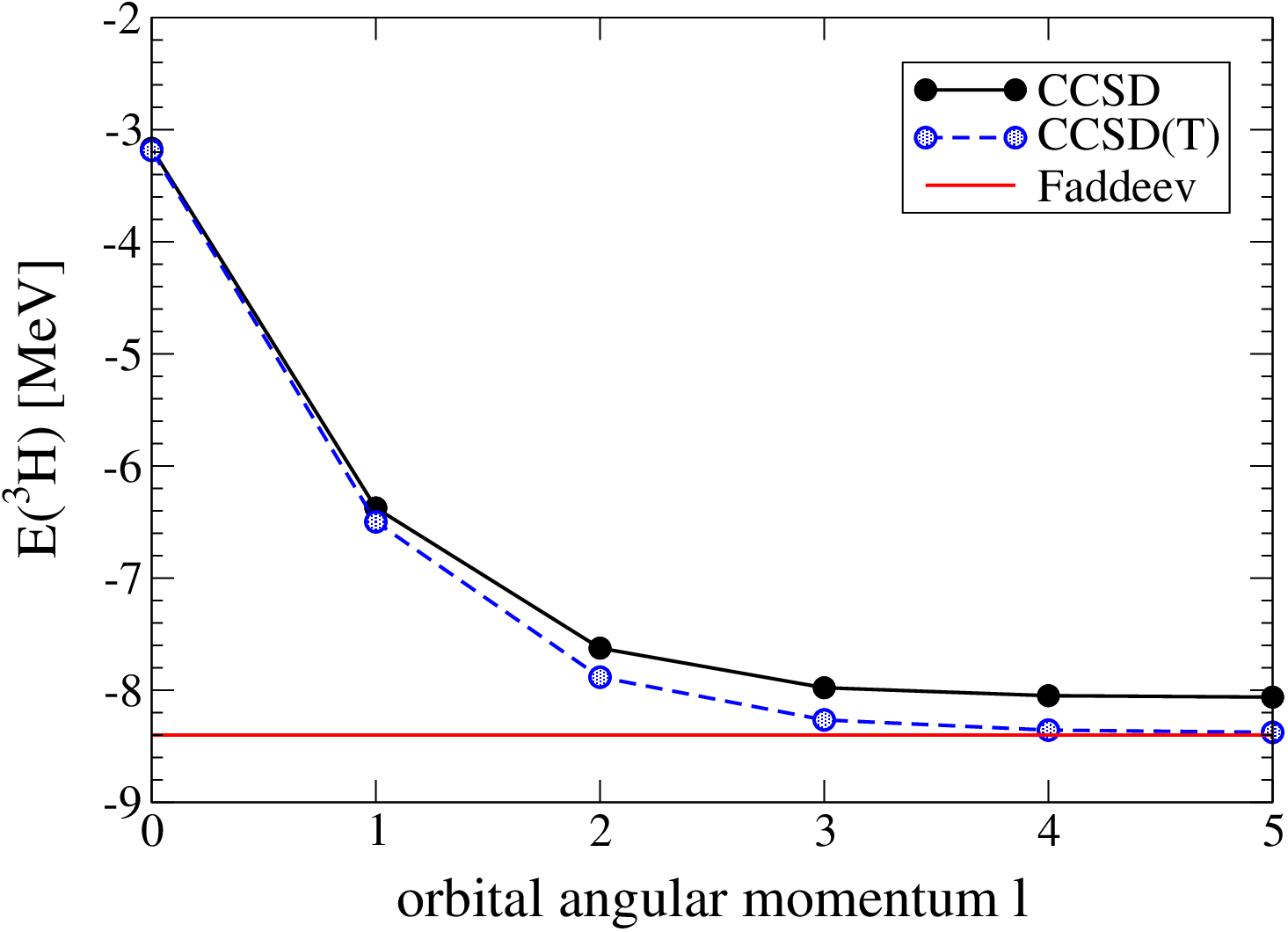}
\caption{(Color online) CCSD and CCSD(T) energies for $^3$H using 
a model space with fixed maximum $N=2n + l=12$ and fixed $\hw=14 \mev$ as 
a function of the maximum orbital angular momentum $l$. For comparison,
we also show the exact Faddeev result of Ref.~\cite{Vlowk3N}.}
\label{fig1}
\end{figure}

\begin{figure}[t]
\includegraphics[width=0.45\textwidth,clip=]{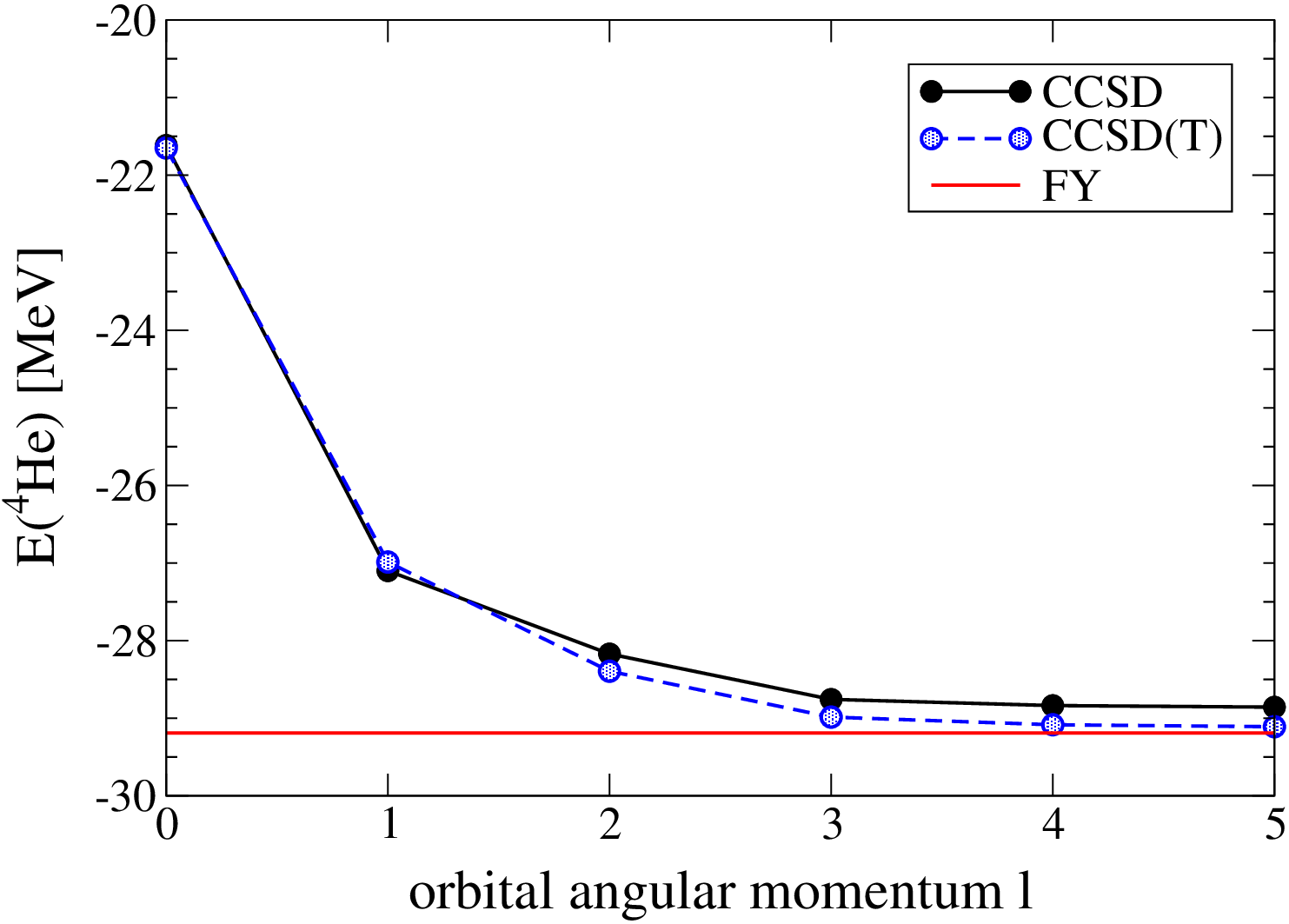}
\caption{(Color online) CCSD and CCSD(T) energies for $^4$He as 
a function of the maximum orbital angular momentum $l$ and
the exact Faddeev-Yakubovsky (FY) result of Ref.~\cite{Vlowk3N}.
For details, see the caption to Fig.~\ref{fig1}.}
\label{fig11}
\end{figure}

In this section, we present our coupled-cluster calculations for the
ground-state energies of $^3$H and $^4$He, and we compare our results
to the exact Faddeev and Faddeev-Yakubovsky energies of
Ref.~\cite{Vlowk3N}. We first discuss in detail the dependence on the
size of the model space and the single-particle basis.  The
coupled-cluster calculations initially used a single-particle basis of
oscillator states whose principal and angular momentum quantum numbers
$n$ and $l$ obey $2n+l \leqslant N$, so $N+1$ is the number of major
oscillator shells included.  Note that in previous
calculations~\cite{Hei99,Mi00,Dean04,Kow04}
$N$ denoted the number of major oscillator shells. However, we
observed that the convergence with respect to the angular momentum $l$
is much quicker, since only low partial waves contribute to low-energy
properties, while the convergence with respect to the principal
quantum number $n$ is slower. This slower convergence is due to the
sharp momentum cutoff used for $\vlowk$. It is intuitively clear that
a harmonic-oscillator representation of an interaction with a sharp
cutoff needs a considerable number of radial wave functions to be
accurate.  The recent work of Refs.~\cite{variational2,smooth}
confirms this picture and demonstrates that smooth cutoffs improve the
convergence in few-body calculations.

Figures~\ref{fig1} and~\ref{fig11} show the convergence of our CCSD
and CCSD(T) energies for $^3$H and $^4$He using a model space with
fixed $N=2n + l=12$ and fixed $\hw=14 \mev$ as a function of the
maximum orbital angular momentum $l$. This implies that for $l=0$ we
include oscillator functions with $n \leqslant 6$ nodes; for $l
\leqslant 1$ we include oscillator functions with $n \leqslant 6$ for
the $s$ states ($l=0$) and $n \leqslant 5$ for the $p$ states ($l=1$),
and so on.  Clearly, the angular momentum quantum number needs not to
exceed $l=5$ for the ground-state energies of $s$-shell
nuclei. Therefore, we limit our single-particle basis to $l \leqslant
5$ for the following coupled-cluster calculations of $^3$H and $^4$He.

\begin{figure}[t]
\includegraphics[width=0.45\textwidth,clip=]{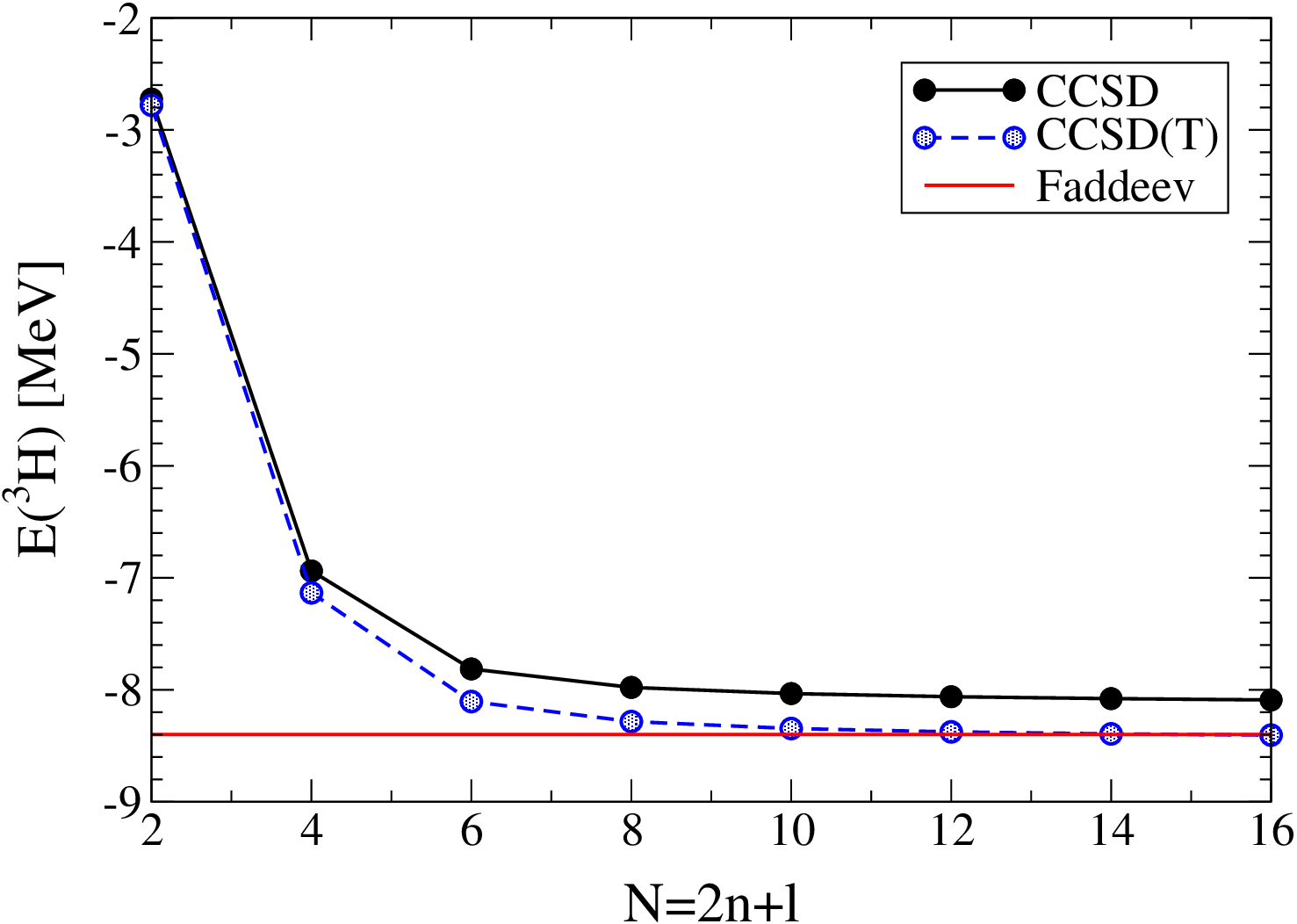}
\caption{(Color online) CCSD and CCSD(T) results for the ground-state 
energy of $^3$H as a function of the model-space size $N=2n+l$, with
fixed $l \leqslant 5$ and fixed $\hw=14 \mev$. For comparison,
we also show the exact Faddeev result of Ref.~\cite{Vlowk3N}.}
\label{fig2}
\end{figure}

\begin{figure}[t]
\includegraphics[width=0.45\textwidth,clip=]{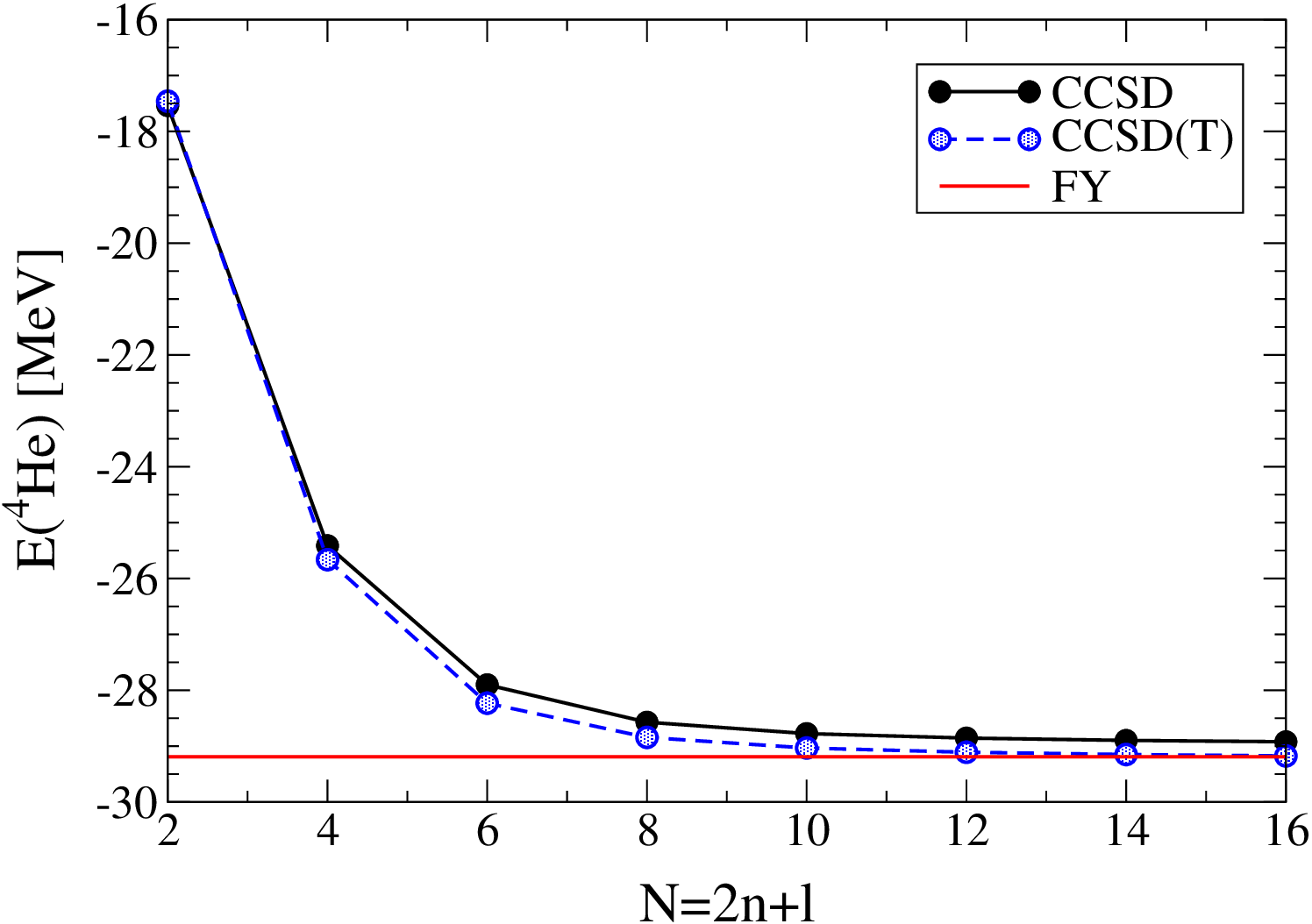}
\caption{(Color online) CCSD and CCSD(T) results for the ground-state 
energy of $^4$He as a function of the model-space size $N=2n+l$ and 
the exact Faddeev-Yakubovsky (FY) result of Ref.~\cite{Vlowk3N}.
For details, see the caption to Fig.~\ref{fig2}.}
\label{fig22}
\end{figure}

In Figs.~\ref{fig2} and~\ref{fig22}, we present our CCSD and CCSD(T) 
results for the ground-state energies of $^3$H and $^4$He as a function
of the model-space size $N$, with fixed $l \leqslant 5$ and fixed
$\hw=14 \mev$. Both CCSD and CCSD(T) energies converge with respect
to the model space size for $N \approx 12 \ldots 14$. 
For the largest model space 
with $N=16$, we obtain for $^3$H,
\begin{eqnarray}
E_{\rm CCSD}(^3{\rm H}) &=& -8.09 \mev \,, \nonumber \\
E_{\rm CCSD(T)}(^3{\rm H}) &=& -8.40 \mev \,, \nonumber
\end{eqnarray}
and for $^4$He,
\begin{eqnarray}
E_{\rm CCSD}(^4{\rm He}) &=& -28.92 \mev \,, \nonumber \\
E_{\rm CCSD(T)}(^4{\rm He}) &=& -29.18 \mev  \,. \nonumber
\end{eqnarray}
The CCSD(T) energies are within $70 \kev$ and $10 \kev$ of the Faddeev and 
Faddeev-Yakubovsky (FY) results~\cite{Vlowk3N} $E(^3{\rm H}) = -8.470(2) 
\mev$ and $E(^4{\rm He}) = -29.19(5) \mev$.

\begin{figure}[t]
\includegraphics[width=0.45\textwidth,clip=]{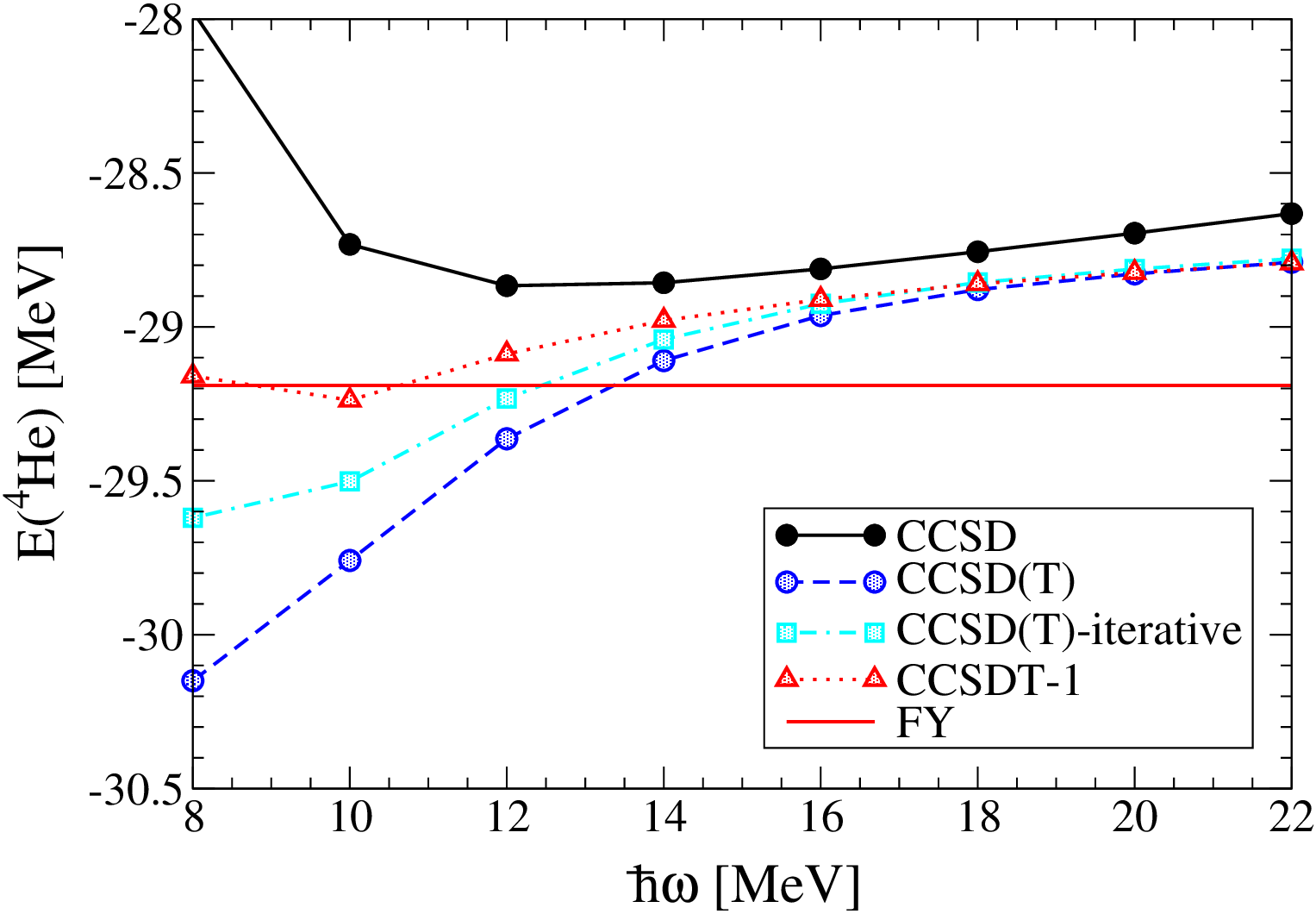}
\caption{(Color online) CCSD and various approximate CCSDT energies
for $^4$He as a function of the oscillator frequency $\hw$, for fixed
$N = 12$ and $l \leqslant 5$. For comparison, we also show the exact
Faddeev-Yakubovsky (FY) result.}
\label{fig3}
\end{figure}

Finally, we study the dependence of our results on the oscillator
frequency $\hw$. This is shown in Fig.~\ref{fig3} for fixed $N = 12$ 
and $l \leqslant 5$. While the CCSD results exhibit a very small 
variation over the shown $\hw$ range, the variation of the perturbative
triples corrections CCSD(T) is somewhat larger. Moreover, the downward
trend of the CCSD(T) energies with decreasing $\hw$ indicates that
perturbative
triples corrections are starting to break down for smaller values of
$\hw$. The non-iterative perturbative triples correction assumes that
we work in a basis where the Fock matrix is diagonal. However, our 
oscillator basis does not diagonalize the Fock matrix, so strict 
calculations would have to iterate triples corrections until 
self-consistency is reached. From Fig.~\ref{fig3} we observe that
iterative CCSD(T) improves on the non-iterative CCSD(T) results,
but also has a downward trend with decreasing $\hw$. Finally,
we present calculations based on the iterative CCSDT-1 approximation
to full CCSDT. CCSDT-1 includes all diagrams through fourth order
in perturbation theory, but contrary to the perturbative CCSD(T)
corrections, the CCSDT-1 approximation is treated self-consistently
and the singles and doubles amplitudes are modified by the triples
amplitude according to Eqs.~(\ref{ccsdt11}) and (\ref{ccsdt12}).
We clearly find that CCSDT-1 improves
on the triples corrections and leads to a very weak dependence on
$\hw$. Note that these CCSDT-1 results are also the first step towards full
CCSDT calculations in nuclear physics.

Our results for the light nuclei $^3$H and $^4$He demonstrate that
coupled-cluster theory meets the benchmarks set by exact methods.
It is therefore interesting to use
this method to establish benchmark energies for heavier nuclei
that other {\it ab-initio} approaches can compare to. This is the
subject of the next section.

\section{Results for $^{16}$O and $^{40}$Ca}
\label{res2}

\begin{figure}[t]
\includegraphics[width=0.45\textwidth,clip=0]{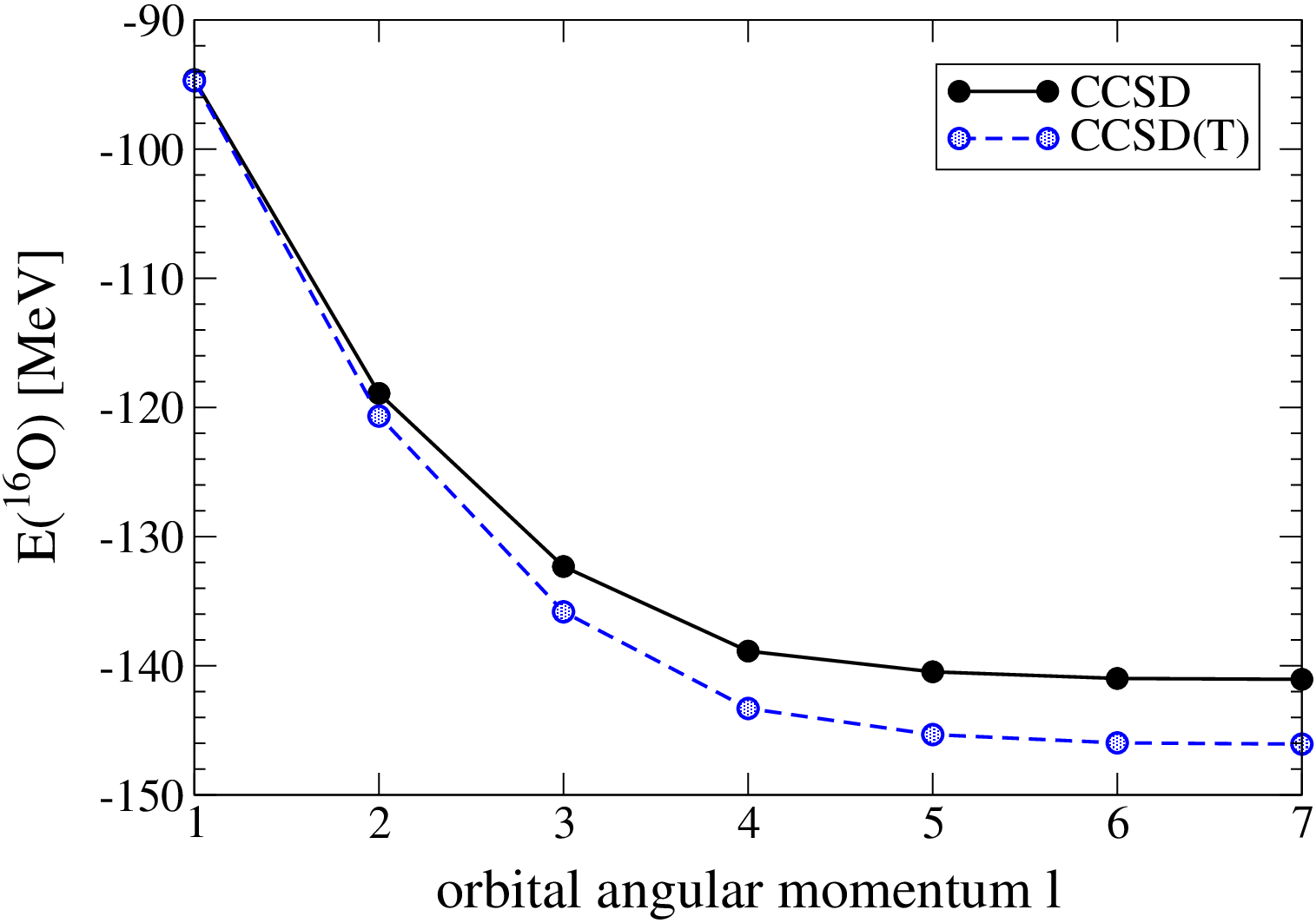}
\caption{(Color online) CCSD and CCSD(T) results for the binding
energy of $^{16}$O using a model space with fixed maximum $N=2n+l=10$
and fixed $\hw=20 \mev$ as a function of the maximum orbital angular 
momentum $l$.}
\label{fig4}
\end{figure}

Next, we present our coupled-cluster calculations for $^{16}$O and 
$^{40}$Ca. In Fig.~\ref{fig4}, we show the convergence of the
CCSD and CCSD(T) results for the binding energy of $^{16}$O using 
a model space with fixed $N=2n+l=10$ and fixed $\hw=20 \mev$ as a 
function of the maximum orbital angular momentum $l$. For $^{16}$O, 
we find that $l \leqslant 7$ is sufficient to reach convergence 
at the $10 \kev$ level. Therefore, we restrict the following
coupled-cluster calculations for $^{16}$O to $l \leqslant 7$.

\begin{figure}[t]
\includegraphics[width=0.45\textwidth,clip=]{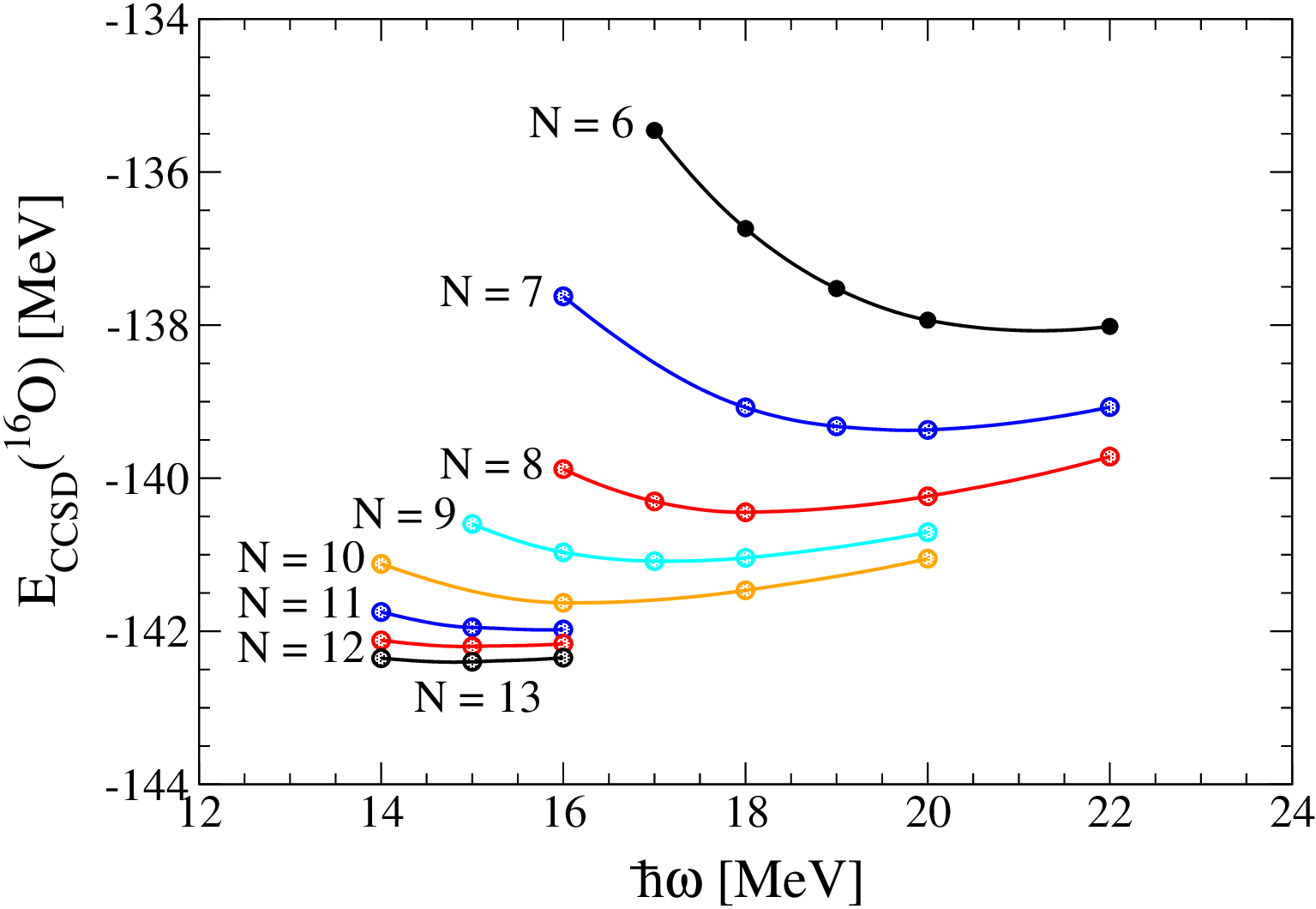}
\caption{(Color online) CCSD results for the binding energy of $^{16}$O
as a function of the oscillator frequency $\hw$ for increasing sizes of
the model space $N=2n+l$ with fixed $l \leqslant 7$.}
\label{fig5}
\end{figure}

\begin{figure}[t]
\includegraphics[width=0.45\textwidth,clip=]{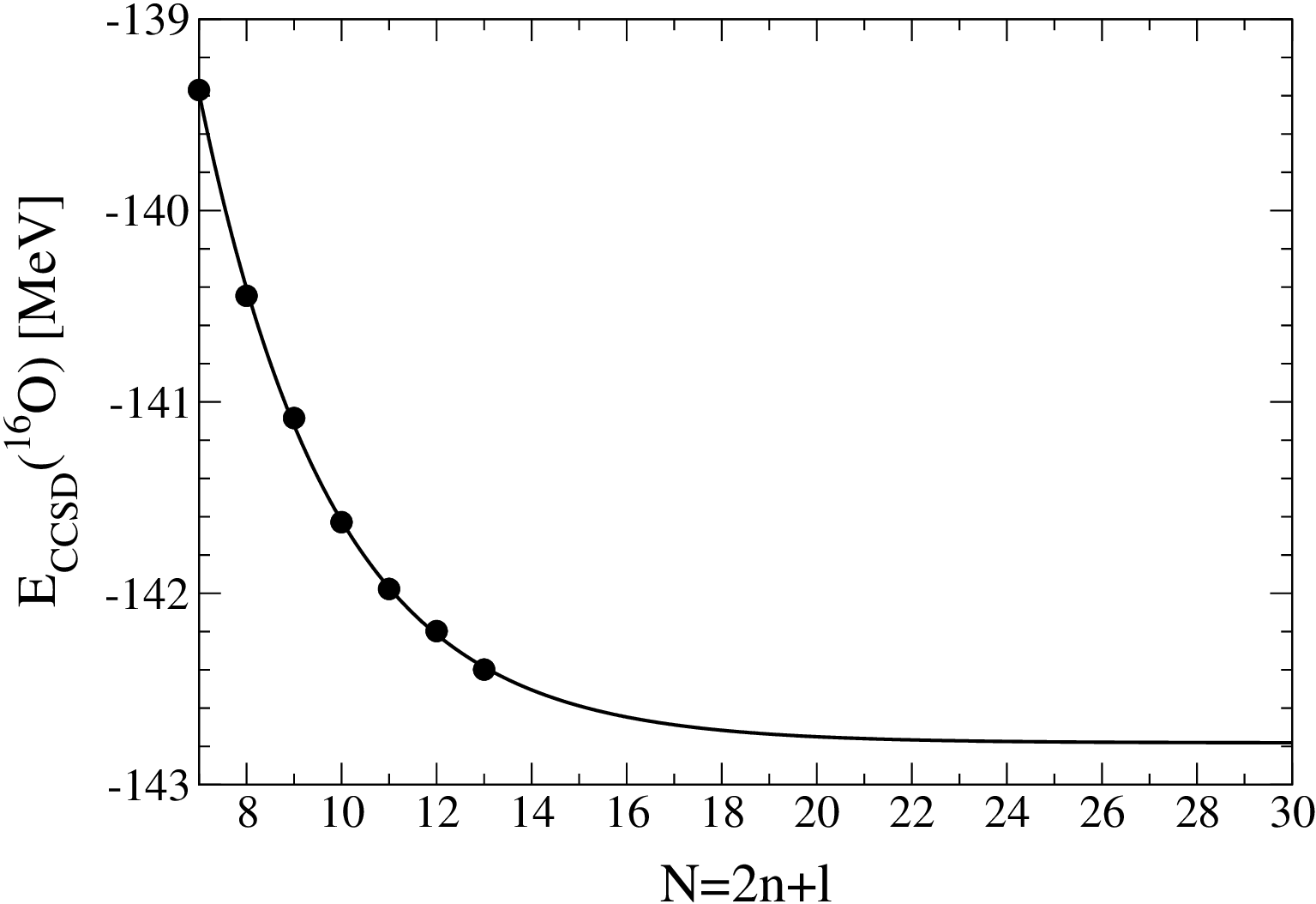}
\caption{(Color online) CCSD results (taken at the $\hw$ minima) for
the binding energy of $^{16}$O as a function of the model-space size 
$N=2n+l$ at fixed $l \leqslant 7$ and exponential fit (solid line).}
\label{fig6}
\end{figure}

Figure~\ref{fig5} shows the dependence of the CCSD binding energies
of $^{16}$O on the oscillator frequency $\hw$ for increasing sizes of
the model space $N$ with fixed $l \leqslant 7$. The largest
calculations for $N=13$ include more than $10^3$ single-particle
orbitals. We observe that the CCSD energies are converged at the
$0.5 \mev$ level and can be used to extrapolate to infinite model
space. This is demonstrated in Fig.~\ref{fig6} where we give the
CCSD energies (taken at the $\hw$ minima) as a function 
of the model-space size $N$ at fixed $l \leqslant 7$.
Using the CCSD minima, we make an exponential fit of the form
$E(N)=E_\infty + a \exp{(-b N)}$ to the data points. The result is
also shown in Fig.~\ref{fig6} and yields the extrapolated infinite
model space value $E_{{\rm CCSD},\infty}(^{16}{\rm O}) = -142.78 \mev$. Our 
largest $N=13$ result is $E_{\rm CCSD}(^{16}{\rm O}) = -142.40
\mev$. The conservative error estimate due to the finite size of the model 
space is thus about 0.5 MeV. 

\begin{figure}[t]
\includegraphics[width=0.45\textwidth,clip=]{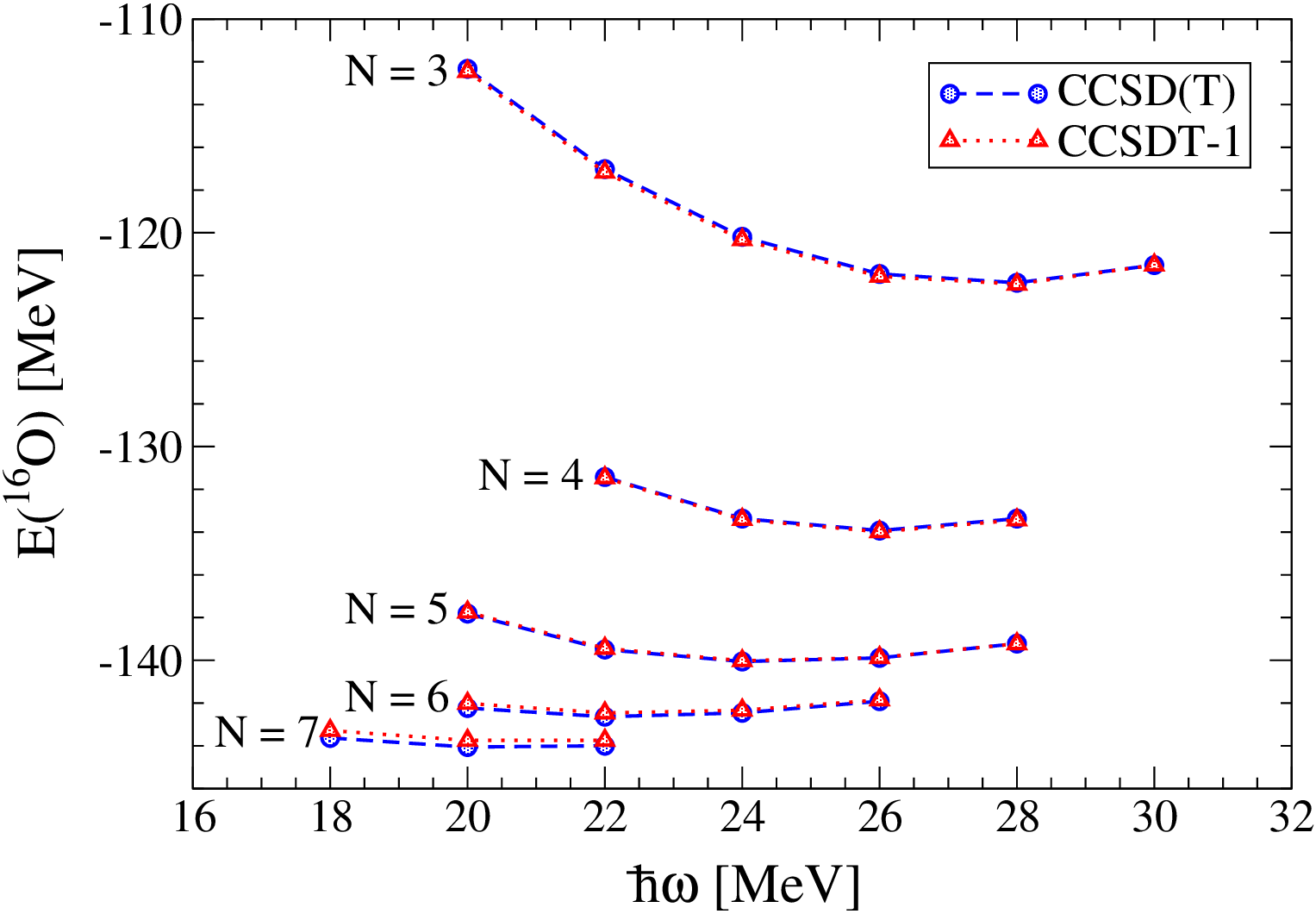}
\caption{(Color online) CCSD(T) and CCSDT-1 results for the binding 
energy of $^{16}$O as a function of the oscillator frequency $\hw$ 
for increasing sizes of the model space $N=2n+l$ ($l \leqslant 7$).}
\label{fig7}
\end{figure}

\begin{figure}[t]
\includegraphics[width=0.45\textwidth,clip=]{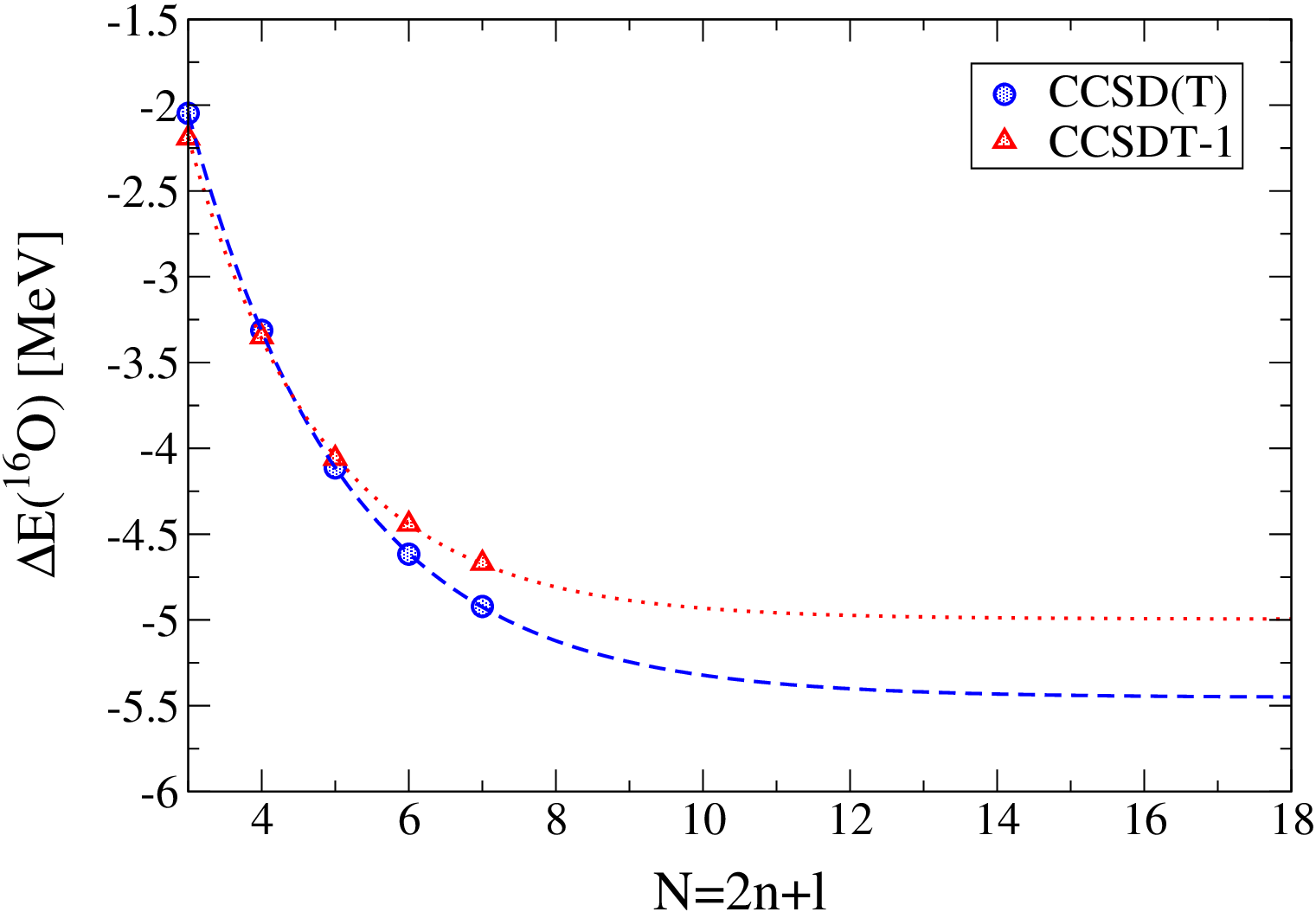}
\caption{(Color online) Contributions to the binding energy of 
$^{16}$O from triples corrections CCSD(T) and CCSDT-1
as a function of the model-space size $N=2n+l$ at fixed $\hw = 22 \mev$
and $l \leqslant 7$. The dashed and dotted lines are exponential fits
and yield the extrapolated energy corrections.}
\label{fig77}
\end{figure}

In Fig.~\ref{fig7}, we study triples corrections to the binding energy
of $^{16}$O via the CCSD(T) and CCSDT-1 approaches as a function of
the oscillator frequency $\hw$ for increasing sizes of the model space
$N$ ($l \leqslant 7$). We present results up to $N=7$ (eight major
oscillator shells), which was the largest model space we could handle
for the CCSDT-1 scheme due to storage limitations. The CCSD(T) and
CCSDT-1 energies agree nicely for the range of oscillator frequencies
and model spaces considered. The only difference is that, as for
$^4$He, the CCSD(T) approach gives slightly more binding than CCSDT-1
for the largest model spaces. The close agreement between CCSD(T) and
CCSDT-1 gives us confidence that the perturbative triples corrections
work well over this regime. 

Let us study the contributions of the triples amplitudes to the
binding energy of $^{16}$O in more detail.  Figure~\ref{fig77} shows
the energy differences $\Delta E = E - E_{\rm CCSD}$ that are due to
triples corrections ``(T)'' and ``T-1'' as a function of the
model-space size $N$ at fixed $\hw = 22 \mev$ and $l \leqslant 7$. The
corresponding exponential extrapolations to infinite model spaces
yield $-5.45 \mev$ from the ``(T)'' correction and $-5.00 \mev$ from
the ``T-1'' correction. This suggests that the error estimate for the
CCSD(T) and CCSDT-1 calculations is about $0.5 \mev$.  This can be
viewed as an error estimate due to the truncation of the cluster
operator.  Combined with the 0.5 MeV uncertainty due to the size of
the model space, we thus arrive at an error estimate of about 1
MeV for $^{16}$O.

We note that $^{16}$O is overbound by about $20 \mev$ when compared to
the experimental binding energy. A similar result has been found by
Fujii {\it et al.}~\cite{Fujii}.  However, a comparison of results
based only on NN interactions to experiment is meaningless, since 3NF
are crucial to describe few- and many-body observables (see for
instance the discussion in Refs.~\cite{Vlowk3N,Vlowknuc,Hag07}). In
nuclear matter, the corresponding 3NF contribution is repulsive and
the expectation values remain consistent with chiral effective field
theory power-counting estimates~\cite{Vlowknuc}.

\begin{figure}[t]
\includegraphics[width=0.45\textwidth,clip=]{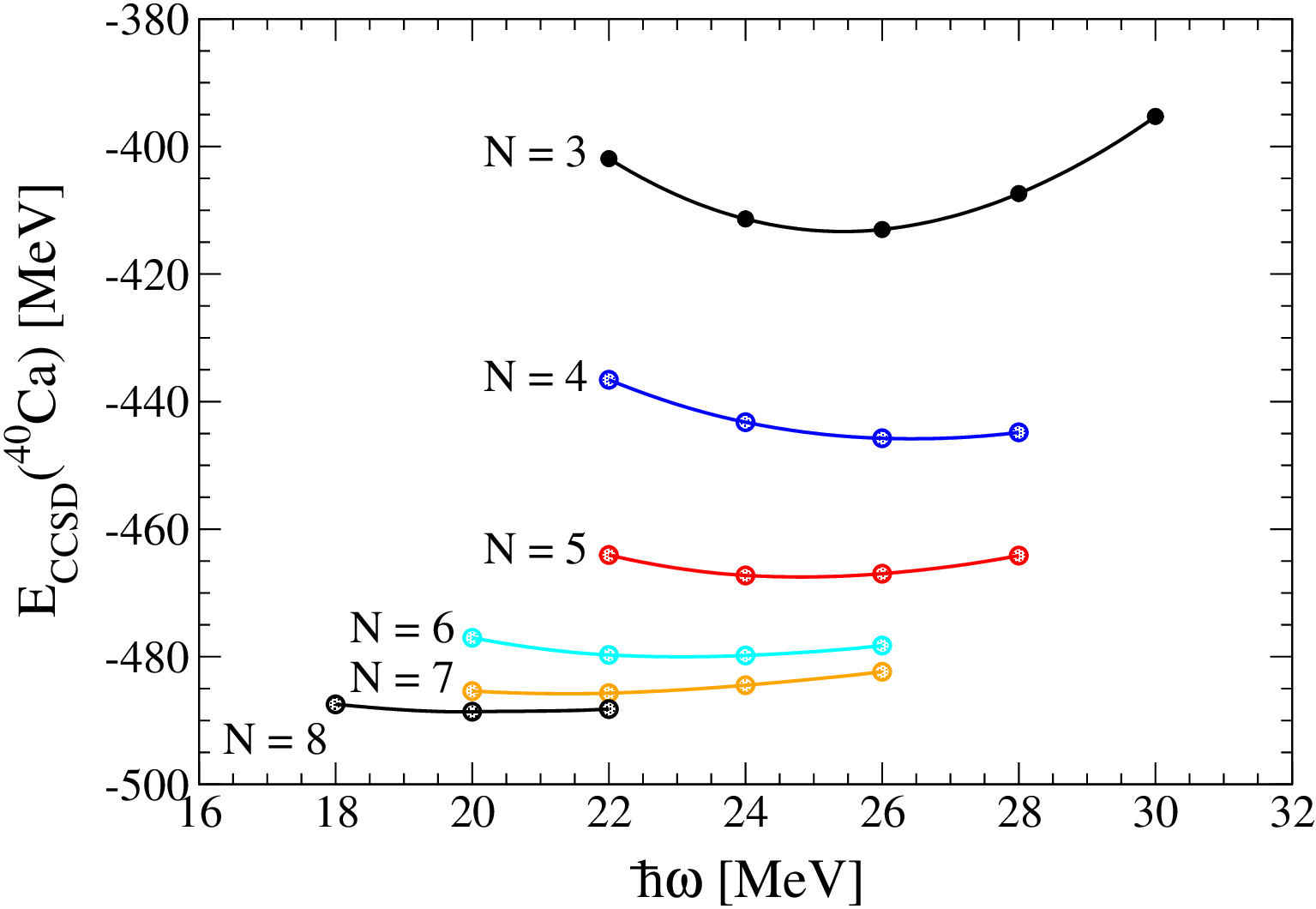}
\caption{(Color online) CCSD results for the binding energy of $^{40}$Ca
as a function of the oscillator frequency $\hw$ for increasing sizes of
the model space $N=2n+l$. (Note that there is no restriction on $l$ for
these model spaces.)}
\label{fig8}
\end{figure}

\begin{figure}[t]
\includegraphics[width=0.45\textwidth,clip=]{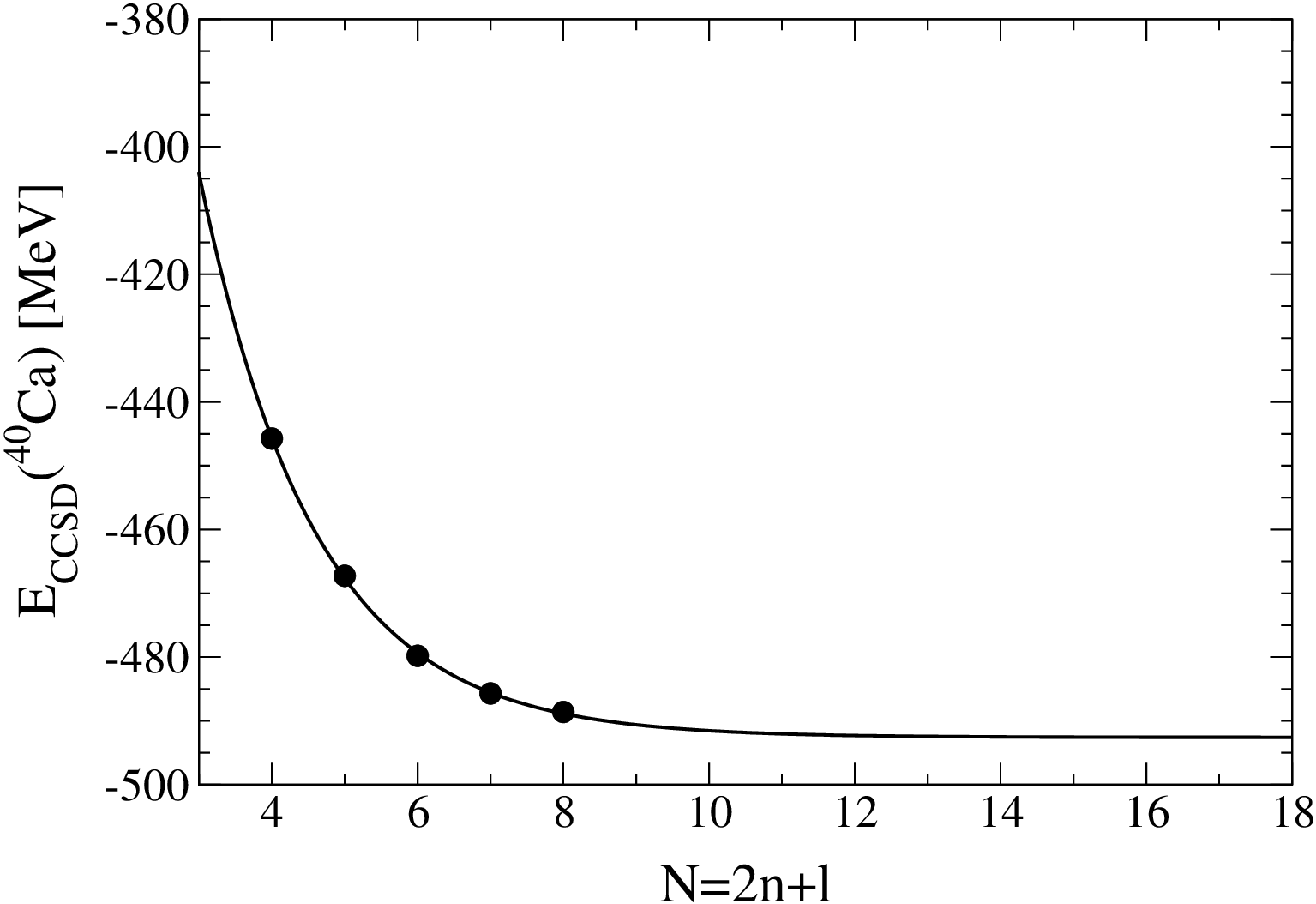}
\caption{(Color online) CCSD results (taken at the $\hw$ minima) for
the binding energy of $^{40}$Ca as a function of the model-space size 
$N=2n+l$ (without restriction in $l$) and exponential fit (solid line).}
\label{fig9}
\end{figure}

Finally, we turn to the more challenging case of $^{40}$Ca.
Figure~\ref{fig8} shows the CCSD binding energy of $^{40}$Ca as a 
function of the oscillator frequency $\hw$ for model spaces up to
$N=8$ (nine major oscillator shells). 
(Note that there is no restriction on $l$ for these model spaces.)
This represents the largest coupled-cluster calculation to date in
nuclear physics. In these largest calculations, we have $40$ active
particles in $660$ single-particle orbitals. The effective shell-model
dimension in this space would be of the order of $10^{63}$.
From Fig.~\ref{fig8}, we find that the CCSD energies of $^{40}$Ca
are converging reasonably well. We again expect that low-momentum 
interactions with smooth cutoffs will lead to even improved convergence.
In Fig.~\ref{fig9}, we present the CCSD energies for $^{40}$Ca (taken 
at the $\hw$ minima) as a function of the model-space size $N$.
The exponential extrapolation to infinite model space yields
$E_{{\rm CCSD},\infty}(^{40}{\rm Ca}) = -492.6 \mev$, and we find
that the CCSD energy for $N=8$ is about 4 MeV from the fully converged
CCSD result.

\begin{figure}[t]
\includegraphics[width=0.45\textwidth,clip=]{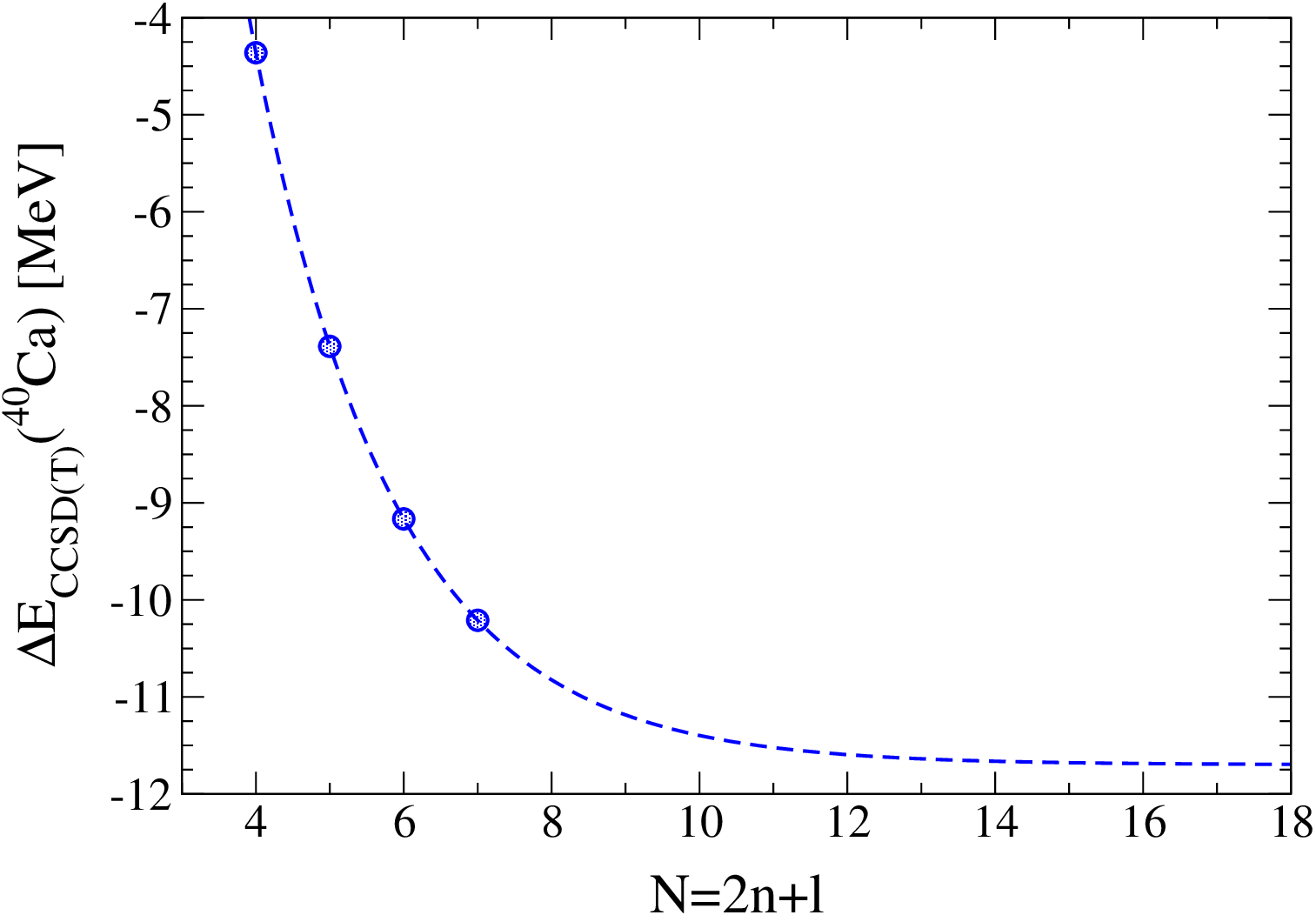}
\caption{(Color online) Contributions to the binding energy of 
$^{40}$Ca from triples corrections CCSD(T) as a function of the 
model-space size $N=2n+l$ at fixed $\hw = 22 \mev$ (without 
restriction in $l$) and exponential fit (dashed line).}
\label{fig99}
\end{figure}

We then perform CCSD(T) calculations and show in Fig.~\ref{fig99} the
triples energy corrections as a function of the model-space size $N$
at fixed $\hw = 22 \mev$. Due to memory limitations, we were not able
to perform CCSDT-1 calculations in model spaces reaching up to $N=7$.
The exponential extrapolation to infinite model space yields $-11.70
\mev$, while the largest $N=7$ result is $-10.21 \mev$.  The
convergence we find with respect to the size of the model space is
similar to the recent results by Fujii {\it et al.}~\cite{Fujii2}.
Recall that the different triples corrections for $^{16}$O differed by
about 10\% from each other. Thus, we estimate that the error due to
the truncation of the cluster operator is about 1 MeV for $^{40}$Ca.
The total error estimate for $^{40}$Ca is thus about 5 MeV, and is
dominated by the uncertainty due to the finite size of the model
space.
 
We summarize our coupled-cluster results for the binding energies 
of $^4$He, $^{16}$O, and $^{40}$Ca in Table~\ref{tab:corr_energy},
which gives the extrapolated correlation energies $\Delta E_{\rm CCSD}$
and $\Delta E_{\rm CCSD(T)}$. We find that for $^4$He, $^{16}$O, and 
$^{40}$Ca, the triples corrections are a factor of $\approx 0.015, 
0.066$, and $0.081$ smaller than the CCSD correlation energies. From
this, we again estimate the missing correlation energy from quadruples, 
pentuplets, and so on, to be of the order of $1 \mev$ for $^{40}$Ca.
We note that $^{16}$O is overbound by about $20 \mev$ and $^{40}$Ca
by about $150 \mev$ when compared to the experimental binding energies.
This is not surprising and points to the importance of 3NF for
nuclear structure calculations~\cite{Vlowk3N,Vlowknuc,Hag07}. 

\begin{table}[t]
\begin{tabular}{l|r|r|r}
\hline
\multicolumn{1}{c|}{} & \multicolumn{1}{c|}{$^4$He} &
\multicolumn{1}{c|}{$^{16}$O} & \multicolumn{1}{c}{$^{40}$Ca} \\
\hline \hline
$E_0$ & -11.8 & -60.2 & -347.5 \\
$\Delta E_{\rm CCSD}$ & -17.1 & -82.6 & -143.7 \\
$\Delta E_{\rm CCSD(T)}$ & -0.3 & -5.4 & -11.7 \\
\hline
$E_{\rm CCSD(T)}$ & -29.2 & -148.2 & -502.9 \\
\hline \hline
exact (FY) & -29.19(5) & & \\
\hline
\end{tabular}
\caption{Reference vacuum energies, $E_0$, CCSD and CCSD(T) correlation
energies, $\Delta E_{\rm CCSD}$ and $\Delta E_{\rm CCSD(T)}$, and binding
energies $E_{\rm CCSD(T)}$ for $^4$He, $^{16}$O and $^{40}$Ca. The vacuum
energies, $E_0$, 
are for $\hw = 14 \mev$ in the case of $^4$He and $\hw
= 22 \mev$ for $^{16}$O and $^{40}$Ca.
The CCSD and CCSD(T) energies are the extrapolated infinite model space
results. The exact Faddeev-Yakubovsky result is from Ref.~\cite{Vlowk3N}}
\label{tab:corr_energy}
\end{table}

We can compare the coupled-cluster energies to the recent
importance-truncated NCSM results of Roth and
Navr{\'a}til~\cite{Roth07} which are based on the same $\vlowk$
interaction.  The importance-truncated NCSM combines a particle-hole
truncation scheme ($4p$-$4h$ for $^{16}$O and $3p$-$3h$ for $^{40}$Ca
in Ref.~\cite{Roth07}) with importance sampling of many-body states
based on perturbation theory. The particle-hole truncation scheme
leads to unlinked diagrams and hence is not
size-extensive~\cite{Bar07}. Using an exponential extrapolation, Roth
and Navr{\'a}til~\cite{Roth07} find binding energies $E= -137.75 \mev$
for $^{16}$O and $E= -461.83 \mev$ for $^{40}$Ca at the minimum in
$\hw$. Our coupled-cluster results idicate that the converged energies
are approximately $10 \mev$ and $40 \mev$ lower for $^{16}$O and
$^{40}$Ca, respectively. Note also that CCSD scales computationally
more favorably than a full $4p$-$4h$ calculation, while it already
includes a considerable part of linked $4p$-$4h$
excitations~\cite{Bar07}.

\section{Summary}
\label{sum}

In summary, we have performed {\it ab-initio} coupled-cluster
calculations for $^3$H, $^4$He, $^{16}$O, and $^{40}$Ca based 
on low-momentum interactions $\vlowk$. At the CCSD(T) level,
the ground-state energies for $^3$H and $^4$He are practically
converged with respect to the size of the model-space and exhibit 
a very weak dependence on the oscillator frequency.
The resulting energies are within $70 \kev$ and $10 \kev$
of the corresponding Faddeev and Faddeev-Yakubovsky benchmarks.
For $^{16}$O and $^{40}$Ca, we estimate that the CCSD(T) binding
energies are converged within 1 MeV and 5 MeV, respectively.
Future calculations will include convergence studies for 
low-momentum interactions with smooth cutoffs~\cite{smooth,SRG}
and advancing the 3NF frontier to medium-mass nuclei based on
the findings of Ref.~\cite{Hag07}. Our results
confirm that coupled-cluster theory is a powerful {\it ab-initio}
method that meets and sets benchmarks.

\section*{Acknowledgments}

We thank S.K.~Bogner, R.J.~Furnstahl and A.~Nogga for useful
discussions.  This work was supported by the U.S. Department of Energy
under Contract Nos. \ DE-AC05-00OR22725 with UT-Battelle, LLC (Oak
Ridge National Laboratory), and DE-FC02-07ER41457 (University of
Washington), and under Grant No.\ DE-FG02-96ER40963 (University of
Tennessee), and by the Natural Sciences and Engineering Research
Council of Canada (NSERC). TRIUMF receives federal funding via a
contribution agreement through the National Research Council of
Canada. Computational resources were provided by the National Center
for Computational Sciences at Oak Ridge and the National Energy
Research Scientific Computing Facility.

\end{document}